\documentclass[twocolumn,aps,showpacs,nofootinbib,epsfig]{revtex4}
\usepackage{amssymb}
\usepackage{amsmath}
\usepackage{graphics}
\usepackage{epsfig}
\usepackage[dvips]{color}

\newcommand{\be}{\begin{equation}}
\newcommand{\ee}{\end{equation}}
\newcommand{\bea}{\begin{eqnarray}}
\newcommand{\eea}{\end{eqnarray}}
\newcommand{\beas}{\begin{eqnarray*}}
\newcommand{\eeas}{\end{eqnarray*}}

\newcommand{\nn}{\nonumber\\}

\begin{document}
\title{Three-hadron angular correlations from pQCD at RHIC and LHC}
\author{Alejandro Ayala$^1$, Jamal Jalilian-Marian$^2$, Antonio Ortiz$^1$, Guy
  Pai\'c$^1$, J. Magnin$^3$ and Maria Elena Tejeda-Yeomans$^4$}
\affiliation{$^1$Instituto de Ciencias Nucleares, Universidad Nacional
Aut\'onoma de M\'exico, Apartado Postal 70-543, M\'exico Distrito Federal
04510, Mexico.\\ 
$^2$Department of Natural Sciences, Baruch College, New York, New York 10010,
USA and CUNY Graduate Center, 365 Fifth Avenue, New York, New York 10016,
USA.\\ 
$^3$Centro Brasileiro de Pesquisas F\'{\i}sicas, CBPF, Rua Dr. Xavier Sigaud
150, 22290-180, Rio de Janeiro, Brazil.\\ 
$^4$Departamento de F\'{\i}sica, Universidad de Sonora, Boulevard Luis
Encinas J. y Rosales, Colonia Centro, Hermosillo, Sonora 83000, Mexico.}

\begin{abstract}
We study three-hadron azimuthal angular correlations in high energy
proton-proton and central nucleus-nucleus collisions at RHIC and LHC at
mid-rapidity. 
We use the LO parton matrix elements for $2\rightarrow 3$ processes and
include the effect of parton energy loss in the Quark-Gluon Plasma using the
modified fragmentation function approach. For the case when the produced
hadrons have either same or not too different momenta, we observe two
away side peaks at $2\pi/3$ and $4\pi/3$. 
We consider the dependence of the angular correlations on energy loss
parameters that have been used in studies of single inclusive hadron
production at RHIC. 
Our results on the angular 
dependence of the cross section agree well with preliminary data 
by the PHENIX collaboration. We comment on the possible contribution of
$2\rightarrow 3$ processes to di-hadron angular correlations and how a
comparison of the two processes may help characterize the
plasma further.

\end{abstract}

\maketitle

\section{Introduction}\label{I}

Jet quenching, the energy loss of a parton moving in a Quark-Gluon Plasma
(QGP), is one of the key paradigms emerging from high energy heavy ion
collisions at the Relativistic Heavy Ion Collider (RHIC) and the Large Hadron
Collider (LHC). This energy loss is believed to cause the suppression of
single hadron transverse momentum spectra in Au + Au collisions, as compared
to p + p collisions, as well as the disappearance of the away-side
peak in two-particle azimuthal correlations~\cite{experiments}. Nevertheless,
the fact that many different energy loss models can fit the
observed suppression of single and double inclusive hadrons, produced in high
energy heavy ion collisions, makes it imperative to consider more
exclusive observables in order to shed light on the dynamics of energy
loss. One can ask for example whether the {\it double hump}
structure on the away-side in azimuthal two-particle correlations in Au + Au
collisions~\cite{doublehump} can be studied from looking at more exclusive
channels such as three-hadron correlations. These studies
have shown to provide a powerful tool to distinguish between different
proposed scenarios~\cite{threepart, Ajitanand}.

This question has been addressed in Ref.~\cite{Ayala2} where we showed that
some properties of the two-particle correlations can be understood from
three-hadron production studies. The connection is made by considering the
different path lengths traveled by the two away side partons in $2\rightarrow
3$ versus the away side parton in $2\rightarrow 2$ parton processes. In the
former, one of the two final state partons in the away side travels, on the
average, a larger path length than the other and 
also larger than the away side parton in the latter processes, leading to an
increase in the probability of absorption of this parton due to energy
loss. Since the parton that travels the smaller path length in $2\rightarrow
3$ processes in the away side has on the average, also a smaller path
length to go through than the away side parton in $2\rightarrow 2$ processes,
the parton in 
the former has a smaller chance of loosing energy than its counterpart in the
latter processes. The combination of these two effects effectively amplifies
the production of structures that upon accumulation of signal look like a
double hump or a broader peak in the away side, depending on the
momentum difference between leading and associate particles. 

To look closer at this issue, in this work we consider the three-hadron
production cross sections in p + p and A + A collisions using leading order
(LO) $2\rightarrow 3$ partonic processes. To include the effect of QGP and
parton energy loss on the final particle spectra, we use modified
fragmentation functions. We then investigate the effect
of energy loss on the angular correlations of the three produced hadrons by 
varying the parameters associated with the energy loss suffered by each
parton. We compare our results to the preliminary data on three-hadron
correlation function measured at RHIC and find that the angular dependence of
the correlation function is reproduced correctly by our
formalism. 

The work is organized as follows: in Sec.~II we write the three-hadron
production cross section in p + p and also introduce the modified fragmentation
function which takes parton energy loss into account for A + A collisions. We
show the time-evolved profile of the medium and the average number of
scatterings as functions of the initial points and trajectories on the
transverse plane for central collisions. In Sec.~III we 
investigate the dependence of the three-hadron production cross section at
RHIC and LHC on energy loss parameters. We then compare our results to
preliminary data on the angular correlations of the three produced hadrons and
show that there is a 
good agreement in the kinematic range accessible by our formalism. We conclude
in Sec.~IV by pointing out to the possible contributions of $2\rightarrow 3$
processes to di-hadron correlations and how they may probe different
transverse plane evolution of the hard scatterings within the plasma.

%%%%%%%%%%%%%%%%%%%%%%%%%%%%%%%%%%%%

\section{Three-hadron production}\label{II}

The differential cross section for three-hadron production, with momenta
$h_1,\ h_2,\ h_3$, in p + p collisions in mid-rapidity is given by 

\bea
   &&\left(\frac{d\sigma}{d\theta_2^{{\mbox{\tiny{H}}}}
   d\theta_3^{{\mbox{\tiny{H}}}}dh_1dh_2dh_3}
   \right)^{{\mbox{\tiny{pp}}} \rightarrow {\mbox{\tiny{H$_1$H$_2$H$_3$}}}}
   =
   \frac{1}{h_1h_2h_3}
   \sum_{i,j} \frac{1}{3!}\nonumber\\
   &&\sum_{\stackrel{\mbox{\tiny{permutations}}}
   {k\neq l\neq m}}
   \frac{F(\theta_2,\theta_3)}{8(2\pi)^4}\ 
    \int_0^1dx_1x_1
   f_{i/{\mbox{\tiny{p}}}}(x_1)x_1f_{j/{\mbox{\tiny{p}}}}(x_1)
   \nonumber\\
   &&
   D_{{\mbox{\tiny{P}}}_k/{\mbox{\tiny{H}}}_1}(z_{1k})
   D_{{\mbox{\tiny{P}}}_l/{\mbox{\tiny{H}}}_2}(z_{2l})
   D_{{\mbox{\tiny{P}}}_m/{\mbox{\tiny{H}}}_3}(z_{3m})
   |{\mathcal{M}_{ij\rightarrow {\mbox{\tiny{P$_k$P$_l$P$_m$}}}}}|^2
   \nonumber\\
\label{ppxsechadron}
\eea  
where $f_{i/{\mbox{\tiny{p}}}}$ and $f_{j/{\mbox{\tiny{p}}}}$ are the
distribution functions of partons $i,j$ within the colliding protons. We use
the CTEQ6 parametrization~\cite{CTEQ6}.
${\mathcal{M}_{ij\rightarrow {\mbox{\tiny{P$_k$P$_l$P$_m$}}}}}$ is the
corresponding leading order matrix element describing the process at the
parton level~\cite{Ellis}. 

The sum is over all the possible colliding parton species, over all of
the three possible partons in the final state and
over all the permutations for a given parton in the final state to
become a given hadron. The magnitudes of the final state parton momenta are
determined from momentum conservation at the parton level as functions of the
angles that $\vec{p}_2$ and $\vec{p}_3$
($\theta_2,\ \theta_3$, respectively) make with $\vec{p}_1$  
\bea
   p_1&=&\frac{x_1\sqrt{s}\sin (\theta_3 - \theta_2)}
              {\sin (\theta_3 - \theta_2) + \sin\theta_2 - \sin\theta_3}\nn
   p_2&=&\frac{-x_1\sqrt{s}\sin\theta_3}
              {\sin (\theta_3 - \theta_2) + \sin\theta_2 - \sin\theta_3}\nn
   p_3&=&\frac{x_1\sqrt{s}\sin\theta_2}
              {\sin (\theta_3 - \theta_2) + \sin\theta_2 - \sin\theta_3},
\label{fixedmomenta}
\eea
where $\sqrt{s}$ is the total center of mass energy available for the
collision. $x_1$ is the momentum fraction of the incoming parton in
the projectile. The momentum fraction of the incoming parton in the target is
fixed also by momentum conservation to be $x_1$. The angular dependent
part of the phase space factor is  
\bea
   F(\theta_2,\theta_3)=\frac{-\sin\theta_3\sin\theta_2
                               \sin(\theta_3 - \theta_2)}
   {[\sin (\theta_3 - \theta_2) + \sin\theta_2 - \sin\theta_3]^4}
\label{anglephasespace}
\eea
and $D_{{\mbox{\tiny{P}}}_n/{\mbox{\tiny{H}}}_m}(h_m/p_n)$,
is the fragmentation function of parton P$_n$ to become hadron H$_m$, 
which is a function of the momentum fraction $z_{nm}=h_m/p_n$ ($n,m=1,2,3$). 
We take the fragmentation functions for charged hadrons as given by the KKP 
parametrization~\cite{KKP}. In
each permutation, any given parton can become the leading hadron which, we
take as H$_1$, and the other two the away-side hadrons. We work in the limit
of collinear fragmentation thus, the angles that define the direction of
the away-side hadrons, $\theta_i^{{\mbox{\tiny{H}}}}$ ($i=2,3$), are linearly
related to the the parton angles $\theta_j$ ($j=2,3$). 

To consider the process within a central heavy-ion collision and thus account
for the effects of energy loss, we resort to the model put forward in
Ref.~\cite{Zhang}. The model considers an initial gluon density obtained from
the overlap of two colliding nuclei, each with a Woods-Saxon density
profile. The gluon density of the medium is diluted only due to longitudinal
expansion of the plasma since transverse expansion is neglected.  

The gluon density $\rho_g$ is
related to the nuclear geometry of the produced medium as
\bea
   \rho_g (\tau, \vec{b}, \vec{r}, \hat{n}) &=&  \frac{\tau_0\,
   \rho_0}{\tau}\, \frac{\pi R_A^2}{2 A}\nn
   &\times& \left[ 
   T_A (|\vec{r} + \hat{n} \tau |)
   +T_A (| \vec{b} - \vec{r} - \hat{n}
   \tau |)\right],
\label{meddens}
\eea
where $T_A$ is the nuclear thickness function, $R_A$ is the nuclear radius and
$A$ the atomic number.

We use the modified fragmentation functions  
\begin{widetext}
\bea
   \tilde{D}_{{\mbox{\tiny{P}}}_n/{\mbox{\tiny{H}}}_m}(z_{nm})
   =\left(1-e^{-\langle L/\lambda\rangle}\right)
   \left[\frac{z_{nm}'}{z_{nm}}
   D_{{\mbox{\tiny{P}}}_n/{\mbox{\tiny{H}}}_m}(z_{nm}') + 
   \langle L/\lambda\rangle\frac{z_{nm;g}'}{z_{nm}}
   D_{{\mbox{\tiny{P}}}_n/{\mbox{\tiny{H}}}_m}(z_{nm;g}')\right] 
   + e^{-\langle L/\lambda\rangle}
   D_{{\mbox{\tiny{P}}}_n/{\mbox{\tiny{H}}}_m}(z_{nm}'),
\label{modfrags}
\eea
\end{widetext}
where $z'_{nm}= h_m/(p_n-\Delta E_n)$ is the rescaled momentum fraction,
of hadron H$_m$ originated from the fragmenting parton P$_n$,
$z_{nm;g}'=\langle L/\lambda\rangle (h_m / \Delta E_n)$ is the rescaled
momentum fraction of the radiated gluon, $\Delta E_n$ is the average
radiative parton energy loss 
and $\langle L/\lambda\rangle$ is the average number of scatterings. The
energy loss  $\Delta E_n$ is related to the gluon density of the produced
medium via 
\bea 
   \Delta E_n = \Bigl\langle \frac{d E_n}{dL}\Bigr\rangle_{1d}\
   t(r,\varphi)
\eea
where
\bea
   t(r,\varphi)\equiv\int_{\tau_0}^{\infty} d 
   \tau \frac{\tau - \tau_0}{\tau_0\, \rho_0}\, 
   \rho_g (\tau, \vec{b}, \vec{r}, \hat{n})
\label{delE}
\eea
%%%%%%%%%%%%%%%%%%%%%%%%%%%%%%%%%%%%%%
\begin{figure}[b!] %fig1
{\centering
{\includegraphics[scale=0.42]{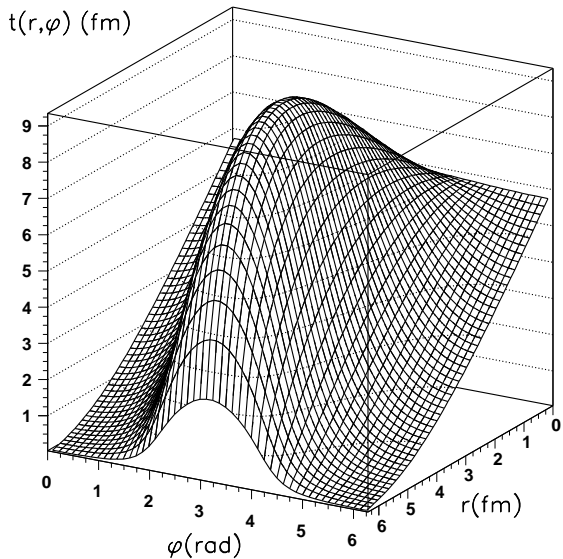}}
}
\caption{Medium influence time $t(r,\varphi )$ for a central collision as a
function of all the initial points and trajectories on the transverse plane
for a central collision. The initial points are characterized by their radial
position $r$ and the angle that the parton trajectory makes with the outward
radial direction $\varphi$.}
\label{fig1}
\end{figure}
%%%%%%%%%%%%%%%%%%%%%%%%%%%%%%%%%%%%%%

\noindent and $ \vec{b}$ is the impact parameter of the collision, 
$\vec{r} $ is the transverse plane location of the hard scattering where the
partons are produced and $\hat{n}$ is the direction in which the produced hard
parton travels in the medium. $r$ and $\varphi$ are the radial position and the
angle that $\hat{n}$ makes with the radial direction, respectively. The
function $t(r,\varphi )$ can be interpreted as the {\it medium influence time},
that is the time during which the medium makes a parton loose energy, when it
was produced at $r$ and starts traveling through the medium making an angle
$\varphi$ with the radial direction. The parameter $\tau_0$ is the formation
time for the medium gluons and $\rho_0$ is the initial central gluon
density. $\tau$ is the time elapsed from the formation time and parametrizes
the path length over the trajectory of the parton 
within the plasma. Since the gluon density profile is a 
rapidly falling function, the upper limit of integration can be safely
set to infinity. Figure~\ref{fig1} shows $t(r,\varphi )$ for a central
collision. Notice that because of the dilution of the medium due to
longitudinal  expansion of the plasma, the biggest effect is on partons which
have to travel less than the full length of the plasma, i.e. partons which are
produced around two thirds from the center of plasma on the transverse plane.

%%%%%%%%%%%%%%%%%%%%%%%%%%%%%%%%%%%%%%
\begin{figure}[b!] %fig2
{\centering
{\includegraphics[scale=0.42]{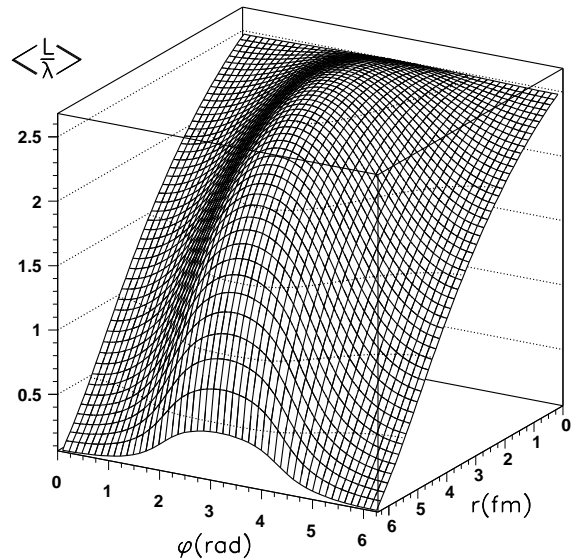}}
}
\caption{Average number of scatterings as a function of all the initial points
and trajectories on the transverse plane for a central collision. The initial
points are characterized by their radial position $r$ and the angle that the
parton trajectory makes with the outward radial direction $\varphi$. Shown is
the case for $\epsilon_0=1.5$ GeV/fm.}
\label{fig2}
\end{figure}
%%%%%%%%%%%%%%%%%%%%%%%%%%%%%%%%%%%%%%
The average number of scatterings for a given parton is 
\bea
   \Bigl\langle\frac{L}{\lambda}\Bigr\rangle &=& \int_{\tau_0}^{\infty}  d
   \tau\frac{1}{\lambda_0\, \rho_0}\, 
   \rho_g (\tau, \vec{b}, \vec{r}, \hat{n}),
\label{avenumscatt}
\eea
where $\lambda_0$ is the parton mean free path. Since we want to consider the
most central collisions, hereafter we set $ \vec{b}=0$. The one dimensional
energy loss $\langle d E_n/d L\rangle_{1d} $ is parameterized as  
\bea
   \Bigl\langle \frac{d E_n}{d L}\Bigr\rangle_{1d} &=& \epsilon_0
   \Bigl[\frac{p_n}{\mu_0} - 
    1.6\Bigr]^{1.2}\Bigl[7.5 + \frac{p_n}{\mu_0}\Bigr]^{- 1}.
\label{dEdL}
\eea
The one-dimensional energy loss per unit length parameter, $\epsilon_0$, is
related to the mean free path $\lambda_0$ by $\epsilon_0\lambda_0=0.5$ GeV.
We work with a value $\mu_0=1.5$ GeV and refer the reader to
Refs.~\cite{Zhang,Gyulassy1,Wang,Salgado} for explicit details of the meaning
and values of the introduced parameters.

Figure~\ref{fig2} shows the average number of scatterings per unit length for
a central collision. Shown is the case for $\epsilon_0=1.5$ GeV/fm for which
the maximum average number of collisions is about 2.5. The average number of
collisions grows with $\epsilon_0$. As is clear from the figure, particles
that experience the largest average number of collisions are the ones that are
produced around one third from the center and travel opposite to the radial
direction, toward the interior of the medium. This effect can also be
understood as arising from the dilution of the medium due to
longitudinal expansion of the plasma.

Given the above, the differential cross section for 
production of three hadrons with momenta $h_1,\ h_2,\ h_3$ at mid-rapidity 
in A + A collisions is given by 
\begin{widetext}
\bea
   \left(\frac{d\sigma}{d\theta_2^{{\mbox{\tiny{H}}}}
   d\theta_3^{{\mbox{\tiny{H}}}}dh_1dh_2dh_3}
   \right)^{{\mbox{\tiny{AA}}} \rightarrow 
   {\mbox{\tiny{H$_1$H$_2$H$_3$}}}}&=&
   \int \frac{d^2r}{\pi R_A^2} 
   \frac{d\varphi}{2\pi}
   \left(\frac{d\tilde{\sigma}}{d\theta_2^{{\mbox{\tiny{H}}}}
   d\theta_3^{{\mbox{\tiny{H}}}}dh_1dh_2dh_3}
   \right)^{{\mbox{\tiny{pp}}} \rightarrow {\mbox{\tiny{H$_1$H$_2$H$_3$}}}},
   \nn
   &=&
   \frac{1}{h_1h_2h_3}
   \sum_{i,j} \frac{1}{3!}
   \sum_{\stackrel{\mbox{\tiny{permutations}}}
   {k\neq l\neq m}}
   \frac{F(\theta_2,\theta_3)}{8(2\pi)^4}\ 
   \int \frac{d^2r}{\pi R_A^2} 
   \frac{d\varphi}{2\pi}
   \int_0^1dx_1x_1
   f_{i/{\mbox{\tiny{p}}}}(x_1)x_1f_{j/{\mbox{\tiny{p}}}}(x_1)
   \nonumber\\
   &\times&
   \tilde{D}_{{\mbox{\tiny{P}}}_k/{\mbox{\tiny{H}}}_1}(z_{1k})
   \tilde{D}_{{\mbox{\tiny{P}}}_l/{\mbox{\tiny{H}}}_2}(z_{2l})
   \tilde{D}_{{\mbox{\tiny{P}}}_m/{\mbox{\tiny{H}}}_3}(z_{3m})
   |{\mathcal{M}_{ij\rightarrow {\mbox{\tiny{P$_k$P$_l$P$_m$}}}}}|^2,
\label{AAxsechadron}
\eea
\end{widetext}
where the new integrations with respect to the p + p case are performed over
the overlap area of the nuclear collision and over the angle $\varphi$ that
the direction of emission of the leading hadron makes with respect to the
radial direction. We have correspondingly divided by the nuclear overlap area
$\pi R_A^2$ and by $2\pi$ in order to have the per nucleon yield. 
%%%%%%%%%%%%%%%%%%%%%%%%%%%%%%%%%%%%%%
%\begin{widetext}
\begin{figure}[b!] %fig3
{\centering
{\includegraphics[scale=0.42]{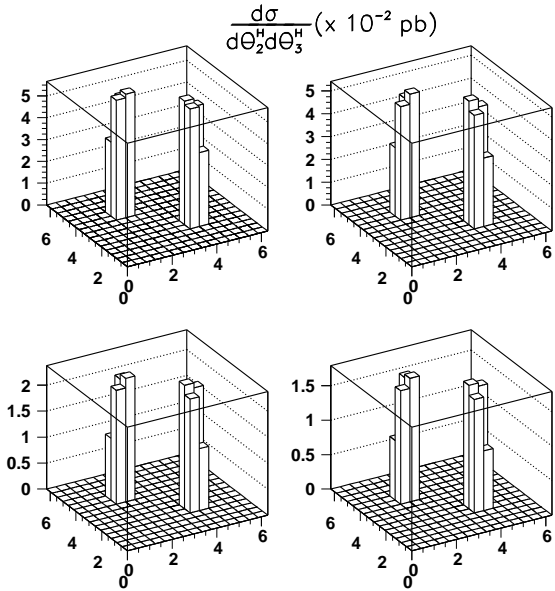}}
}
\caption{
Cross section
($d\sigma/d\theta_2^{{\mbox{\tiny{H}}}}d\theta_3^{{\mbox{\tiny{H}}}}$) for a
symmetric hadron momentum configuration 
where $2$ GeV $\leq h_1,\ h_2,\ h_3 \leq 3$ GeV with
$\sqrt{s_{NN}}=200$ GeV. From left to right and top to bottom are the p + p
case and the A + A cases for $\epsilon_0=0.1,\ 1.5,\ 2$ GeV/fm,
respectively.}
\label{fig3}
\end{figure}
%\end{widetext}
%%%%%%%%%%%%%%%%%%%%%%%%%%%%%%%%%%%%%%

%%%%%%%%%%%%%%%%%%%%%%%%%%%%%%%%%%%%%%
\begin{figure}[b!] %fig4
{\centering
{\includegraphics[scale=0.42]{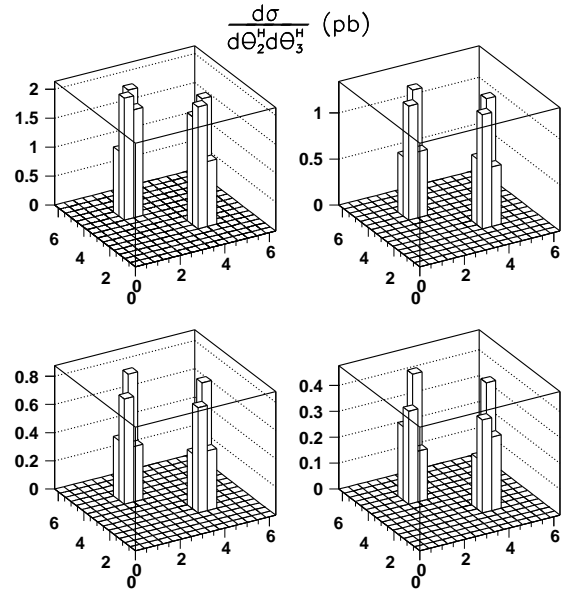}}
}
\caption{
Cross section
($d\sigma/d\theta_2^{{\mbox{\tiny{H}}}}d\theta_3^{{\mbox{\tiny{H}}}}$) for a
symmetric hadron momentum configuration 
where $2$ GeV $\leq h_1,\ h_2,\ h_3 \leq 3$ GeV with
$\sqrt{s_{NN}}=2.76$ TeV. From left to right and top to bottom are the p + p
case and the A + A cases for $\epsilon_0=2,\ 3,\ 4$ GeV/fm,
respectively.}
\label{fig4}
\end{figure}
%%%%%%%%%%%%%%%%%%%%%%%%%%%%%%%%%%%%%%

The matrix elements
${\mathcal{M}_{ij\rightarrow{\mbox{\tiny{P$_k$P$_l$P$_m$}}}}}$ in
Eqs.~(\ref{ppxsechadron}) and~(\ref{AAxsechadron}), representing the
$2\rightarrow 3$ LO QCD hard scattering amplitudes at the parton level are 
divergent due both to collinear and soft singularities. These kinematic 
divergences are universal and can be removed using well known
techniques~\cite{cataniseymour, soper}. Here instead we apply
angular cuts to avoid the divergent regions. An alternative is to  
use an automated version of the dipole subtraction method~\cite{madipole} in
the MadGraph environment~\cite{madgraph}. This avenue is being explored and
will be reported elsewhere. Hereafter,
${\mathcal{M}_{ij\rightarrow{\mbox{\tiny{P$_k$P$_l$P$_m$}}}}}$ stands for the 
finite matrix elements, after applying angular cuts. Throughout we use the
factorization and renormalization scales $\mu_f$ and $\mu_r$ as
$\mu_r=\mu_f=2$ GeV.

%%%%%%%%%%%%%%%%%%%%%%%%%%%%%%%%%%%%%%
\begin{figure}[t!] %fig5
{\centering
{\includegraphics[scale=0.42]{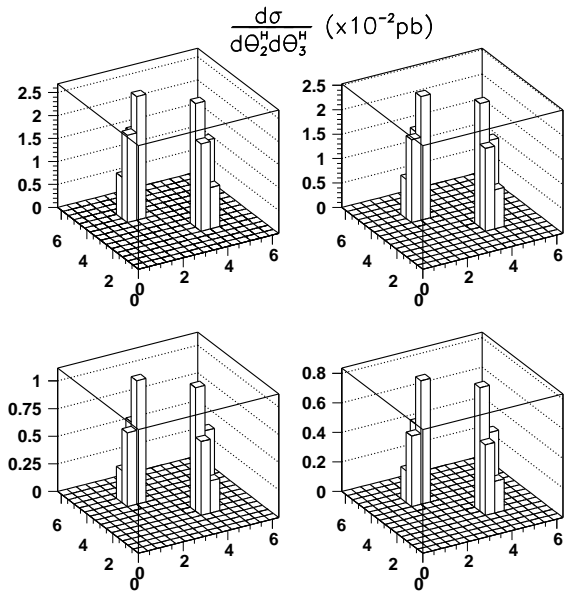}}
}
\caption{
Cross section
($d\sigma/d\theta_2^{{\mbox{\tiny{H}}}}d\theta_3^{{\mbox{\tiny{H}}}}$) for an
asymmetric hadron momentum configuration where $3$ GeV $\leq h_1 \leq 4$ GeV
and 2 GeV $\leq h_2,\ h_3 \leq 3$ GeV with 
$\sqrt{s_{NN}}=200$ GeV. From left to right and top to bottom are the p + p
case and the A + A cases for $\epsilon_0=0.1,\ 1.5,\ 2$ GeV/fm,
respectively.}
\label{fig5}
\end{figure}
%%%%%%%%%%%%%%%%%%%%%%%%%%%%%%%%%%%%%%

\section{Three-hadron correlations}\label{III}

Figure~\ref{fig3} shows the contribution of $2\rightarrow 3$ processes to
the three-hadron correlation function. Shown is the cross section
($d\sigma/d\theta_2^{{\mbox{\tiny{H}}}}d\theta_3^{{\mbox{\tiny{H}}}}$) for a
symmetric hadron momentum configuration 
where $2$ GeV $\leq h_1,\ h_2,\ h_3 \leq 3$ GeV with a collision energy of
$\sqrt{s_{NN}}=200$ GeV, appropriate for RHIC energies. From left to right and
top to bottom, the figure shows the p + p case and the A + A cases for 
values of the one-dimensional energy loss per unit length parameter
$\epsilon_0=0.1,\ 1.5,\ 2$ GeV/fm, respectively. These values are chosen
to correspond to the ones explored in Ref.~\cite{Zhang}. In
these plots
$\pi /5\leq\theta_2^{\mbox{\tiny{H}}} (\theta_3^{\mbox{\tiny{H}}})\leq 4\pi
/5$, $6\pi /5\leq\theta_3^{\mbox{\tiny{H}}}(\theta_2^{\mbox{\tiny{H}}})\leq
9\pi /5$. The excluded region corresponds, 
to events with only two hadrons in the final state, within our resolution. We
also notice that when  
$\epsilon_0\rightarrow 0$, the correlation function in the p + p case is
recovered and that with increasing values of $\epsilon_0$ the intensity of the
signal decreases, as expected.

%%%%%%%%%%%%%%%%%%%%%%%%%%%%%%%%%%%%%%
\begin{figure}[t!] %fig6
{\centering
{\includegraphics[scale=0.42]{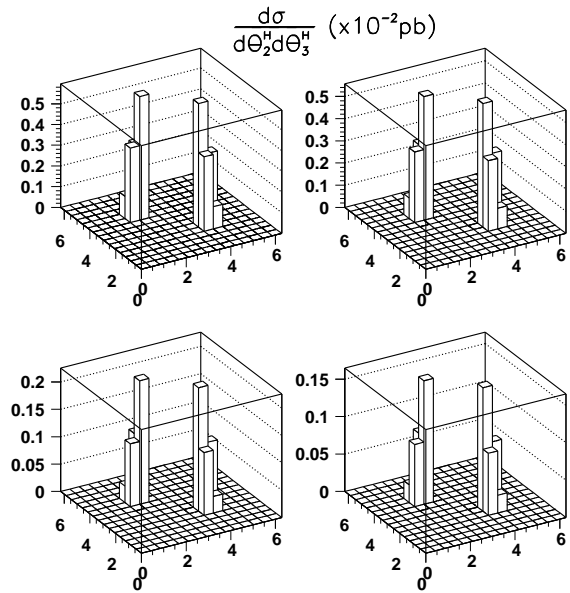}}
}
\caption{
Cross section
($d\sigma/d\theta_2^{{\mbox{\tiny{H}}}}d\theta_3^{{\mbox{\tiny{H}}}}$) for an
asymmetric hadron momentum configuration where $5$ GeV $\leq h_1 \leq 6$ GeV
and 2 GeV $\leq h_2,\ h_3 \leq 3$ GeV with 
$\sqrt{s_{NN}}=200$ GeV. From left to right and top to bottom are the p + p
case and the A + A cases for $\epsilon_0=0.1,\ 1.5,\ 2$ GeV/fm,
respectively.}
\label{fig6}
\end{figure}
%%%%%%%%%%%%%%%%%%%%%%%%%%%%%%%%%%%%%%

Figure~\ref{fig4} shows also the contribution of $2\rightarrow 3$ processes to
the three-hadron correlation function for a symmetric hadron momentum
configuration  $2$ GeV $\leq h_1,\ h_2,\ h_3 \leq 3$ GeV but this time for
a collision energy of $\sqrt{s_{NN}}=2.76$ TeV, appropriate for LHC
energies. From left to 
right and top to bottom, the figure shows the p + p case and the A + A cases
for values of the one-dimensional energy loss per unit length parameter
$\epsilon_0=2,\ 3,\ 4$ GeV/fm, respectively. As in Fig.~\ref{fig3} for
these plots $\pi
/5\leq\theta_2^{\mbox{\tiny{H}}}(\theta_3^{\mbox{\tiny{H}}})\leq 4\pi /5$,
$6\pi /5\leq\theta_3^{\mbox{\tiny{H}}}(\theta_2^{\mbox{\tiny{H}}})\leq 9\pi
/5$. We also notice from 
Fig.~\ref{fig4} that with increasing values of $\epsilon_0$ the intensity of
the signal decreases, as expected.

Figures~\ref{fig5} and~\ref{fig6} show asymmetric hadron momenta
configurations. Figure~\ref{fig5} shows the case where $3$ GeV $\leq h_1 \leq
4$ GeV and 2 GeV $\leq h_2,\ h_3 \leq 3$ GeV, whereas Fig.~\ref{fig5} is the
case where $5$ GeV $\leq h_1 \leq 6$ GeV and 2 GeV $\leq h_2,\ h_3 \leq 3$
GeV, both calculated with  $\sqrt{s_{NN}}=200$ GeV. Notice that the position
of the two peaks on the away side remain at roughly $2\pi/3$ and $4\pi/3$ rad,
as was the case for the symmetric configurations. Their
intensity decreases with respect to the symmetric, lower momentum case of
Fig.~\ref{fig3}, as the difference between away and leading particle momenta
increases but the higher value becomes sharper. Also, the intensity decreases
as the energy loss parameter increases. 

\begin{figure}[t!] %fig7
{\centering
{\includegraphics[scale=0.42]{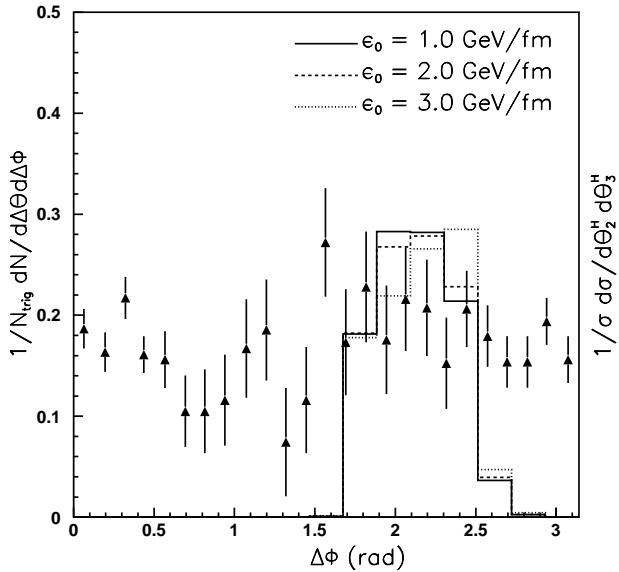}}
}
\caption{Cross section $d\sigma
  /d\theta_2^{\mbox{\tiny{H}}}\theta_3^{\mbox{\tiny{H}}}$ as a function of the 
  angular difference $\Delta\phi = \theta_3^{\mbox{\tiny{H}}} -
  \theta_2^{\mbox{\tiny{H}}}$ for a leading hadron momentum $2.5$ GeV $\leq
  h_1 \leq 4$ GeV and away-side hadron momenta $1$ GeV $\leq h_2,\ h_3 \leq
  2.5$ GeV integrated over a  
  $\Delta\theta=\theta_2^{\mbox{\tiny{H}}}$ angular range $1.65 \leq
  \Delta\theta \leq 2.2$ rad. The histograms are normalized to their cross
  section, $\sigma$, obtained by integration of the differential cross section
  over the above angular ranges and correspond to three values of the energy
  loss parameter $\epsilon_0=1, 2, 3$ GeV/fm. The calculation is compared to
  preliminary data from PHENIX.}  
\label{fig7}
\end{figure}

In order to extract information from the three-hadron correlation function in
a two dimensional analysis, one possibility is to look at this object as a
function of the angular difference $\Delta\phi = \theta_3^{\mbox{\tiny{H}}} -
\theta_2^{\mbox{\tiny{H}}}$ of the away side particles for a range of angles
of one of the away side-particles, say 
$\Delta\theta = \theta_2^{\mbox{\tiny{H}}}$. Figure~\ref{fig7}   
shows this correlation in a A + A environment with $\sqrt{s_{NN}}=200$ GeV for
a leading hadron momentum $2.5$ GeV $\leq h_1 \leq 4$ GeV and away-side hadron
momenta $1$ GeV $\leq h_2,\ h_3 \leq 2.5$ GeV, integrated over a
$\theta_2^{\mbox{\tiny{H}}}$ angular range $1.65 \leq
\Delta\theta \leq 2.2$ rad. Shown are the histograms obtained for
$\epsilon_0=1, 2, 3$ GeV/fm normalized to their cross section, $\sigma$,
obtained by integration of the differential cross section over the above
angular ranges and  compared to preliminary data from
PHENIX~\cite{Ajitanand}. Due to our angular cuts, meant to exclude collinear
hadron production, the regions $\Delta\phi\simeq 0,\ \pi$ rad are not
accessible. However, as is clear from the figure, the angular
region $1.5\lesssim\Delta\phi\lesssim 2.7$ rad is well described,
particularly for the value of $\epsilon_0=2$ GeV/fm.

\section{Conclusion} \label{IV}

%%%%%%%%%%%%%%%%%%%%%%%%%%%%%%%%%%%%%%
%%%%%%%%%%%%%%%%%%%%%%%%%%%%%%%%%%%%%%

In this work we have calculated the three-hadron production cross sections in
both p + p and A + A collisions and investigated the effect of parton energy
loss on the angular correlations between the produced hadrons. We have
considered collision energies appropriate for RHIC and LHC. For the cases
where the hadrons have same or not too 
different momenta, studied in this work, we observe two peaks on the away
side, with respect to the leading hadron, at roughly $2\pi/3$ and $4\pi/3$
rad. The location of the peaks stays the same and their
intensity decreases as the difference between away and leading particle momenta
increases. The intensity also drops as we increase the energy loss parameter. 

There are a few interesting questions that warrant further work, for instance,
one could consider the contribution of these $2\rightarrow 3$ processes
to di-hadron correlations. This could happen when
either one of the partons travels a large distance in the plasma and loses a
large portion of its energy, so that its transverse momentum is outside the
transverse 
momentum window considered for the associate hadrons. Another possibility
comes from the projection
of the three-hadron angular correlation onto two-dimensions,  
i.e., when one integrates over one of the away side hadrons' angle. 

It would be interesting to compare these contributions with those of genuine
$2\rightarrow 2$ processes. Naively, one might expect that for the case when
there are three hadrons in the final state, the partonic hard scattering must
happen close to the center of the plasma in the transverse plane. On the other
hand, one expects that contributions of $2\rightarrow 3$ processes to di-hadron
correlations come from events where the partonic hard scattering 
happens closer to the edge, so that at least one of the partons
travels a large distance through the medium and has a high probability
of being completely absorbed. A study of the transverse plane location
dependence of these events may thus provide more information for the
tomography of the Quark-Gluon Plasma than is possible with single or even
genuine double inclusive hadron production. This work is in progress and will
be reported elsewhere. 

\section*{Acknowledgments}

A.A., J.J-M. and M.E.T-Y. thank CBPF for their kind hospitality and support
during a visit where this work was completed.
Support for this work has been received in part by CONACyT (Mexico) under
grant number 128534, PAPIIT-UNAM under grant number IN103811-3, {\it Programa
de Intercambio UNAM-UNISON}, the DOE Office of Nuclear Physics
through Grant No.\ DE-FG02-09ER41620, by the ``Lab Directed Research
and Development'' grant LDRD~10-043 (Brookhaven National Laboratory),
and by The City University of New York through the PSC-CUNY Research
Award Program, grant 63404-0041.

\end{document}